\documentclass[journal=jctcce,manuscript=article]{achemso}

\usepackage[version=3]{mhchem} 
\usepackage{hyperref}

\author{Samuel Mathews}
\author{Xiaodan Zhu}
\author{Andr\'e Guerra}
\author{Phillip Servio}
\author{Alejandro Rey}
\email{alejandro.rey@mcgill.ca}
\affiliation[McGill University]
{Department of Chemical Engineering, McGill University, Montreal, QC H3A 0C5, Canada}

\title{Atomistic Modeling of Methane and Carbon Dioxide Structure I Gas Hydrates Under Pressure: Guest Effects and Properties}

\begin{document}

\begin{tocentry}

%
%
%

\includegraphics[width=9cm]{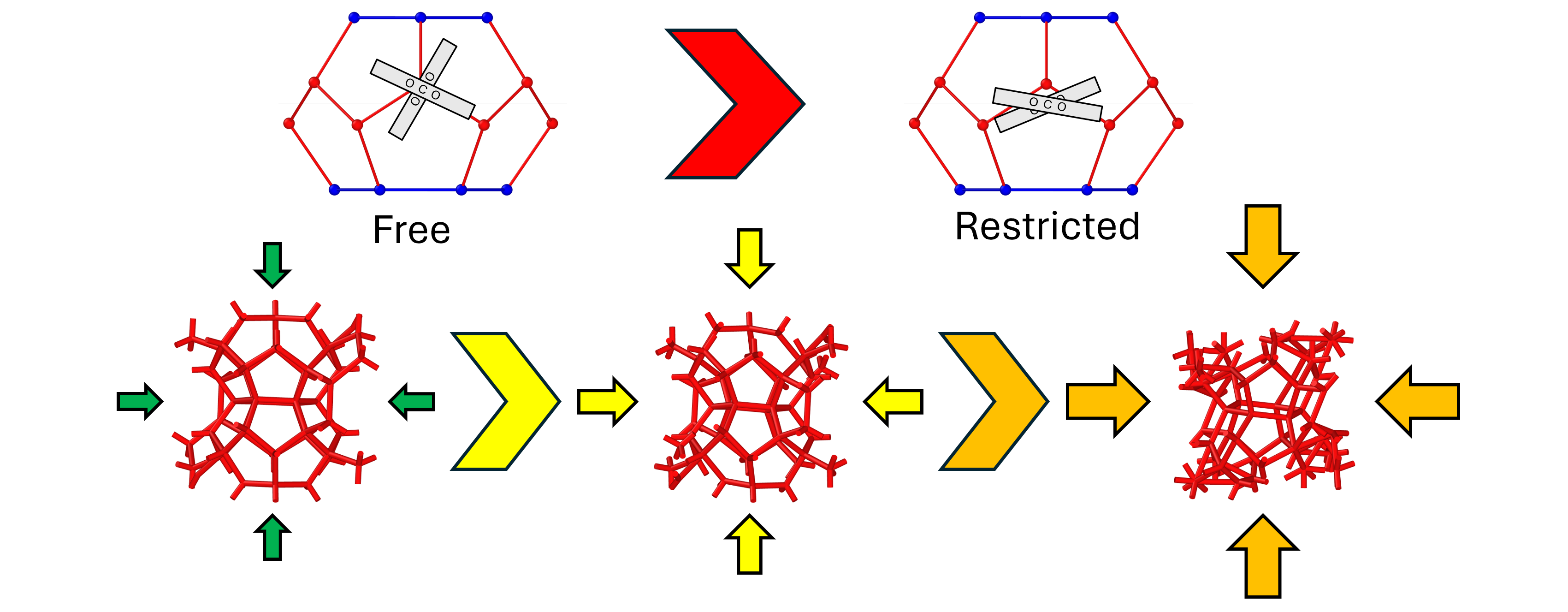}

\end{tocentry}

\noindent\textbf{Preprint / Submitted Manuscript Notice.}
This document is the authors' original submitted manuscript (a preprint) and has not been peer reviewed.
A peer-reviewed version of this work was subsequently published in \textit{Journal of Chemical Theory and Computation}
(copyright \textcopyright\ 2026 American Chemical Society).
For the final edited and formatted version (Version of Record), see:
\textit{J.\ Chem.\ Theory Comput.}\ 2026, 22, 6, 3114--3124.
\href{https://doi.org/10.1021/acs.jctc.5c01868}{\texttt{10.1021/acs.jctc.5c01868}}.

\begin{abstract}
  Gas hydrates are potential candidates in future energy sources while simultaneously providing structures with extensive applications in carbon capture and storage, gas transport, and important separation processes. Prior research in the field considers the dynamics of the water molecule backbone in particular. We investigated the pressure-enthalpy landscape and mechanical stability envelope of sI methane and carbon dioxide hydrates simulated using DFT. We investigated the effect of the revPBE + DFT-D2 and the SCAN + rVV10 and their treatment of the exchange correlation interactions. We examined the zero pressure material properties, finding that revPBE comparatively underbinds the interactions, causing more flexible structures with large equilibrium volumes. Under pressure, the carbon dioxide molecule was found to align itself parallel to the hexagonal faces of the large cage despite the functional used. Additionally, the property differences are caused by the ability of the carbon dioxide molecule to rotate and disperse the changes in the energy landscape in ways that methane molecules cannot. This computational methodology describes the elastic stability of gas hydrate, marginal stability, and critical differences across important molecular interactions, confirming experimentally observed restrictions in guest molecule rotations and novel pressure behaviors under hydrostatic loads.
\end{abstract}

\section{Introduction}

When gaseous molecules and water come into contact at low temperatures and high pressures, solid inclusion compounds called gas hydrates (or clathrate hydrates) may be formed.
Hydrogen-bonded water molecules form cages encapsulating the gas forming a crystalline lattice stabilized by the guest-host interactions.
Initial focus on gas hydrates revolved around their formation in, and plugging of, gas tranmission piplines, where the presence of natural gas, water, and favorable thermodynamic conditions may lead to their nucleation and agglomeration~\cite{Koh2011FundamentalsApplicationsGas}.
Their ability to capture molecules led to further research into their structure, as their unique packaing abilities and structural makeup yield interesting properties and applications in carbon capture and storage, gas storage, desalination, and separation processes~\cite{Cheng2020ReviewProspectsHydrate}.
Additionally, the combination of their storage capacity and formation conditions precipitated analysis of continental shelves that led to estimates suggesting that hydrates in these locations may contain nearly double the carbon content of current global fossil fuel reserves combined~\cite{Demirbas2010MethaneGasHydrate,Carroll2014NaturalGasHydrates}.
The multitude of situations either promoting hydrate formation or actively seeking to reduce their nucleaiton and growth leads to the requirement for multiscale characterization of these phenomena.

The two hydrates encountered in the petroleum industry are structure I (sI) and structure II (sII), with a third less common type being structure H (sH).
The temperature and pressure conditions are an important part of dictating which structure is formed, but the size and type of guest molecule is the strongest influence.
sI and sII hydrates possess cubic unit cells, while sH types have hexagonal unit cells.
Small gas molecules like methane, carbon dioxide, or ethane may for sI hydrates, where 46 water molceules form six tetrakaidecahedrons ($5^{12}6^{2}$) and two pentagonal dodecahedrons ($5^{12}$). 
When larger guest molecules like propane and isobutane, or mixtures of methane and ethane, are involved, sII hydrates form, made of 136 water molecules arranged into eight hexakaidecahedrons ($5^{12}6^{4}$) and 16 $5^{12}$ cages.
sH hydrates, in turn, form when 36 water molecules encapsulate combinations of small and large guests like methane and neohexane into two irregular dodecahedrons ($4^{3}5^{6}6^{3}$), three $5^{12}$ cages, and one much larger icosahedron ($5^{12}6^{8}$)~\cite{Mathews2022RecentAdvancesDensity}.
Once formed, changing thermodynamic conditions may also lead to transitions from one structure to another, depending on the location in the phase space~\cite{Ballard2006OptimizingThermodynamicParameters}.
The structural and guest differences lead to some differences of varying magnitude among their properties, like thermal diffusivity, heat capacity, thermal expansion, and bulk modulus.
Additionally, they share some structural similarities to hexagonal ice (ice I$_h$), leading to improper use of its properties as substitutes for those of hydrates~\cite{Koh2011FundamentalsApplicationsGas}.

Investigating the mechanical and thermodynamic material properties of gas hydrates, such bulk modulus, work of compression, density, and pressure-volume relationships is of great importance in the aforementioned applications as they govern the stability of these materials under varying load conditions. 
Obtaining these properties is often difficult in experimental settings because they depend on cage occupancy, guest molecule identity, and control over the specific structure being formed.
Controlling each of these accuately is difficult and can influence other complementary measurements~\cite{Carroll2014NaturalGasHydrates}.
This leads to frequent discrepancies in property values reported in literature~\cite{Ghafari2019ThermalPropertiesBinary}.
As such, an opportunity arises to perform initial examinations of gas hydrates using simulations, such as density function theory (DFT) and molecular dynamics (MD), to elucidate the molecular mechanisms at the root of macroscale behavior and to improve transferability of analysis from one structure to the next, and to other hydrogen-bonded crystalline structures~\cite{Mathews2022RecentAdvancesDensity}.

Ab initio DFT provides a way to predict and rationalize properties without relying on empirical parameters fitted to experimental measurements, as would be the case in a top-down approach to material modeling, where structures are produced without control at the atomic length scale.
Conversely, DFT uses a bottom-up approach, allowing researchers to identify and develop fundamental theories regarding materials in a reliable and flexible way~\cite{Giustino2014MaterialsModellingUsing}.
Using this approach, we can investigate the macroscopic property changes atom-by-atom, inducing desired properties or studying complex structures from their simplest underlying forms~\cite{Ghavanloo2019FundamentalTenetsNanomechanics}.
Additionally, DFT can reduce issues existing in employing empirical models, where complex and large systems either require too many parameters to accurately describe or too many restrictions are placed on regions of application, as it relies mainly on experimental measurements of the atomic positions to perform electronic structures calculations.
In practice, however, DFT calculations involve calculations that are only exact in principal, and depend on accurate modeling of a critical energetic contribution, the exchange-correlation (XC) energy.
Approximations of this energy are computed with the XC functional, and the choice of which functional to employ is a critical design decision in all subsequent computations~\cite{Rappoport2009ApproximateDensityFunctionals}.

In this work, we investigate the material properties of sI gas hydrates that encapsulate methane and carbon dioxide guest molecules, as well as determine the effect of hydrostatic pressure on these two structures.
We determine how resistant these structures are to compression, as well as what structure is favored under pressure, a critical measure for applications in methane-carbon dioxide exchange for carbon capture and storage~\cite{Hanssens2024EnablingLowcostDecentralized}.
Additionally, we compare the effect of the choice of XC functional on the material properties to show that this choice is a critical parameter in predictive analysis of the the structures.
We extend our previous work~\cite{Mathews2025GeometricCharacterizationsNonUniform} by focusing on the changes of guest molecule behaviors under pressure and under different functionals, instead of focusing on the water backbone structure, to examine the atomic root of fracture mechanics, and differentiate the behaviors.
We employ DFT, as implemented in the Vienna ab initio Simulation Package (VASP)~\cite{Kresse1996EfficiencyAbinitioTotal,Kresse1996EfficientIterativeSchemes,Kresse1993InitioMolecularDynamics,Kresse1999UltrasoftPseudopotentialsProjector} to determine the ground state structure of the hydrate at zero Kelvin, and then compress or stretch this structure while examining the material property changes and how the guest molecules participate in this dynamic.

The organization of this paper is as follows.
First, we describe the computational techniques employed for this work and the important choices that are made in calculations.
Then, we discuss the choice of equation of state (EOS) for determination of material properties for the sI hydrate structures, as well as examine differences depending on the XC functional. 
Next, we study the systems at different pressures, corresponding to the tensile stability limit and the compressive stability limit, thoroughly describing the convergence behaviors, enthalpy differences, bulk moduli, and guest molecule movements.
Finally, we discuss the implications of these findings for the stability of sI methane and carbon dioxide hydrates, the choice of XC functional in future theoretical works, and how the guest molecule may interact with the water lattice during engineering applications.

\section{Methodology}

\subsection{General Computational Details}

The computations were performed using first principles DFT as implemented in the VASP code~\cite{Kresse1996EfficiencyAbinitioTotal,Kresse1996EfficientIterativeSchemes,Kresse1993InitioMolecularDynamics,Kresse1999UltrasoftPseudopotentialsProjector}.
We employed the projector augmented wave potentials~\cite{Blochl1994ProjectorAugmentedwaveMethod,Kresse1999UltrasoftPseudopotentialsProjector} for all calculations, with a plane wave energy cutoff of 520 eV, in line with a high accuracy mode in VASP, and an augmentation grid for increased accuracy~\cite{2019PRECVaspwiki}.
The ionic optimization force tolerance was 0.5 meV$\cdot${\AA}$^{-1}$, while the electronic energy tolerance used was 10$^{-6}$ eV. 
These tolerances combined with a k-point mesh of 36 irreducible points ensured proper convergence of the total energy in all calculations and excellent sampling of the Brillouin zone.

To examine the effect of the XC functional on calculated properties, we employed two different ones.
First, we employed the revised Perdew-Burke-Ernzerhof (revPBE) functional~\cite{Zhang1998CommentGeneralizedGradient} and the DFT-D2 dispersion correction method of Grimme~\cite{Grimme2006SemiempiricalGGAtypeDensity} based on its previous performance in studying the material properties of gas hydrates~\cite{Jendi2015IdealStrengthMethane,Daghash2021FirstPrinciplesElasticAnisotropic,Vlasic2016AtomisticModelingStructure,Zhu2022PiezoelasticityStabilityLimits,JafariDaghalianSofla2023AtomisticgeometryInspiredStructurecompositionproperty}.
This generalized gradient approximation (GGA) provides a balance between accuracy and low computational cost when compared to others of this type.
Secondly, we employed the Strongly Constrained and Appropriately Normed (SCAN) XC functional.
This meta-GGA functional provides the best error elimination for atoms and Jellium spheres~\cite{Singh2017StudyAccurateExchangecorrelation} and works well for describing the water structure~\cite{Wang2019FirstPrinciplesCalculationWater}, providing the accuracy of a more computationally expensive hybrid functional at the cost of a GGA functional~\cite{Sun2015StronglyConstrainedAppropriately,Sun2016AccurateFirstprinciplesStructures}.
Combined with the modified version of the Vydrov-van Voorhis (rVV10) nonlocal correlation functional of Peng et al. for long-range dispersion interactions~\cite{Dion2004VanWaalsDensity,Roman-Perez2009EfficientImplementationVan,Peng2016VersatileVanWaals,Lee2010HigheraccuracyVanWaals}, this XC functional has shown promising results in capturing thermal properties of hydrates using DFT~\cite{Mathews2020HeatCapacityThermal}.

SCAN is a semilocal nonempirical functional that satisfies all known possible exact constraints required, while being appropriately normed where possible.
GGAs, in constrast, often are developed in a way that requires them to pick which constraints, among various incompatible ones, to satisfy~\cite{Sun2015StronglyConstrainedAppropriately}.
One important adaption and improvement in meta-GGAs is the addition of the kinetic energy density to the functional, adding the possibility of satisfying both energy and structure constraints.
In particular, SCAN is not built with bonded information but uses the strong constraints and appropriate norms to enable high accuracy for diversely bonded materials in a non-empirical manner, solidifying its first principles capabilities.
Additionally, it is able to provide more accuracy energies than hybrid GGAs (those that mix exact exchange with that from other calculated sources), which fail to describe Van der Waal (VdW) interactions critical in ice and hydrate structures.
SCAN is able to establish the relative stability of ice phases and predict volume changes between said phases in agreement with experimental results~\cite{Sun2016AccurateFirstprinciplesStructures}.

It is also able to predict the intermediate-range VdW interactions in its accurate descriptions of ice polymorphs and the ordering of liquid water~\cite{Sun2013DensityFunctionalsThat,Zhao2006NewLocalDensity}.
It overestimates the strength of hydrogen bonds compare to revPBE, but overall performs better for systems with diverse bonding characteristics~\cite{Sun2016AccurateFirstprinciplesStructures}.
The treatment of VdW interactions with rVV10 adheres to first principles behaviors through its versatility and applicability to systems of dominant and negligible interactions of this type.
The combination of SCAN + rVV10 is versatile and applies to molecular complexes, bulk solids, and challenging layered materials~\cite{Peng2016VersatileVanWaals}.
These benefits and behaviors will be critical in explaining differences in material properties and guest behaviors analyzed herein.
Unless otherwise noted, we refer henceforth to the revPBE + DFT-D2 as revPBE and the SCAN + rVV10 as SCAN for brevity, unless otherwise noted.

A single unit cell of the sI hydrate structure was simulated.
It contains 46 water molecules that accomodate guest molecules in eight cages, one moleculer per cage.
There are six large cages and two small cages in the sI structure~\cite{Sloan2008ClathrateHydratesNatural}.
Initial coordinates of the oxygen atoms were obtained by Takeuchi et al.~\cite{Takeuchi2013WaterProtonConfigurations} by X-ray diffraction and the protons were placed according to the Bernal-Fowler ice rules such that the net dipole moment was zero and the potential energy was the lowest.
Two guests were studied: methane and carbon dioxide.
For both guests, the carbon atom is placed near the center of the cage and the molecule is randomly oriented to ensure that no bias exist for the simulation.
Ionic optimization is performed after the placement to obtain the corresponding grond state configuration for each system.
We only considered structures with 100\% cage occupancy.
Finally, hydrostatic pressure was applied to each unit cell, from $-1.1$ GPa to $7.5$ GPa to encompass the tensile and compressive stability limits of the hydrate~\cite{Zhu2022PiezoelasticityStabilityLimits}, and at each pressure the system is fully relaxed to its ground state for the given conditions.

\subsection{Equation of State for Solids}

Determining the structure-property relationships for solids begins with obtaining an EOS, starting with a cold isothermal condition~\cite{Becke1988DensityfunctionalExchangeenergyApproximation} as a basis, which would then be extended for temperature effects with thermal expansion coefficient corrections accounting for these~\cite{Stacey1977ApplicationsThermodynamicsFundamental}.
Fitting the energy-volume relationship to an EOS has proven to be a successful technique for extracting meaningful structural strength information for sI, sII, and sH hydrates~\cite{Vlasic2016AtomisticModelingStructure,Zhu2022PiezoelasticityStabilityLimits,Daghash2019StructuralPropertiesSH,JafariDaghalianSofla2023AtomisticgeometryInspiredStructurecompositionproperty}.
We characterize this relationship using the Vinet isothermal EOS, repsented in equation~\ref{eqn:vinet}~\cite{Vinet1987TemperatureEffectsUniversal} in its energy explicit form.
The Vinet relation is selected over the Murnaghan\cite{Murnaghan1944CompressibilityMediaExtreme}, the Birch-Murnaghan~\cite{Birch1947FiniteElasticStrain}, and the Liu~\cite{Li2005FourparameterEquationState} because it has been found to be the least sensitive to pressure range over which it is regressed~\cite{Vlasic2016AtomisticModelingStructure} for hydrate structures and it performs best among common EOS when the pressure range is large enough to adequately capture the curvatuve of the energy-volume dataset correctly~\cite{Vocadlo2000GruneisenParametersIsothermal}.
The fitting parameters are the ground state volume ($E_{0}$), the bulk modulus ($B_{0}$), and the pressure derivative of the bulk modulus ($B_{0}^{'}$).

\begin{equation}
  \label{eqn:vinet}
  \begin{gathered}
  \Delta{E}(V) = \frac{4V_{0}B_{0}}{\left(B_{0}^{'} - 1\right)^{2}}
  \left(
    1 - 
    \left(
      1 - \frac{3}{2}\eta
    \right)
    \mathrm{exp}
    \left(
      \frac{3}{2}\eta
    \right)
  \right) \\
  \eta = \left\{B_{0}^{'} - 1\right\}\left\{1 - \left(\frac{V}{V_{0}}\right)^\frac{1}{3}\right\}
  \end{gathered}
\end{equation}

\section{Results and Discussion}

\subsection{Equilibrium and Ground State}

\begin{figure}
  \includegraphics[width=5in]{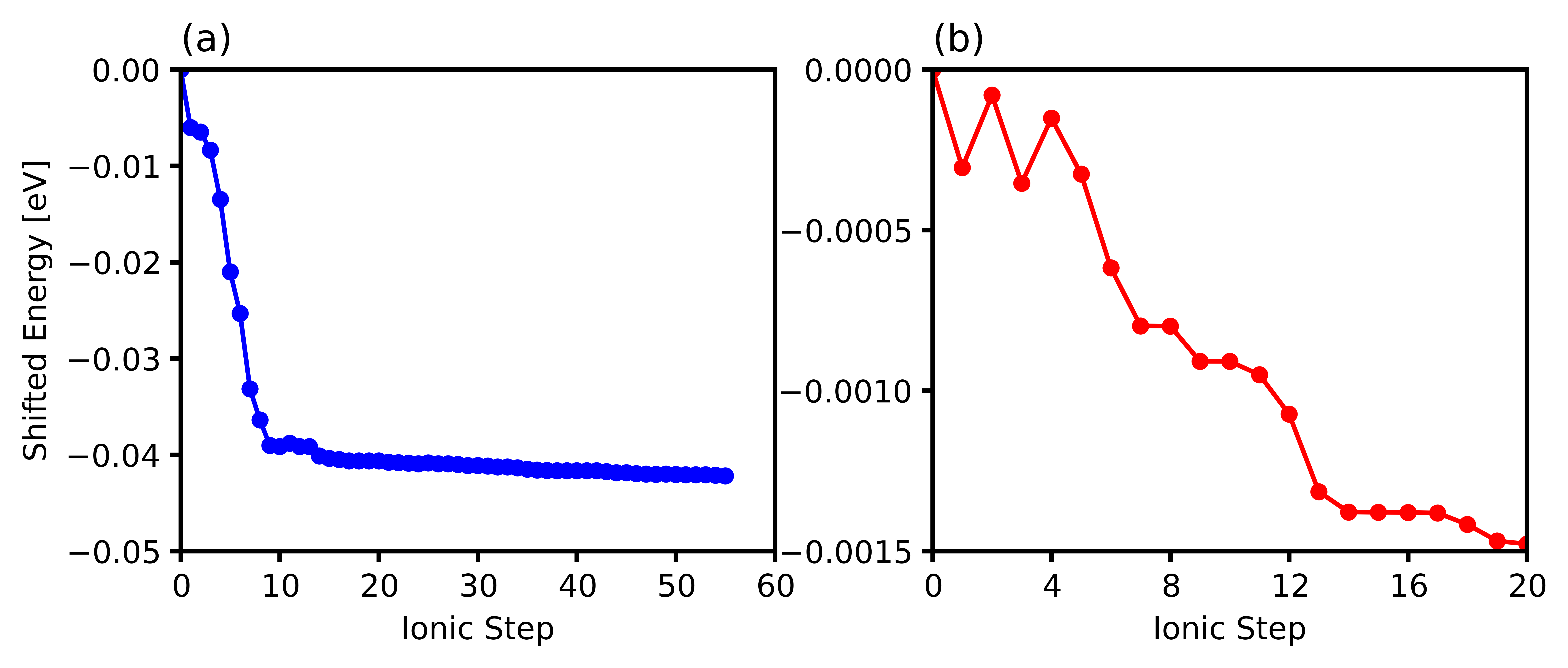}
  \caption{Representative energetic convergence behaviors of the ionic optimization for two sI gas hydrate systems, showing a decrease towards the ground state. (a) The shifted energy as a function of ionic step for the methane sI hydrate calculated using the revPBE XC functional at zero hydrostatic pressure. (b) The shifted energy as a function of ionic step for the carbon dioxide sI hydrate calculated using the SCAN XC functional at 0.4 GPa of hydrostatic pressure.}
  \label{fig:equil}
\end{figure}

In DFT, great care must be taken to ensure that the system in question has indeed reached its ground state configuration~\cite{Ullrich2011ReviewGroundstateDensityfunctional}.
For each pressure, guest molecule, and functional, we ensure that the system has indeed reached the desired tolerance for the electronic self-consistant loop and for the ionic optimization.
Then, we rerun ground state simulations with tighter tolerance to confirm the positions, and to provide a performance balance against running long simulations at very tight tolerances from the outset.
Additionally, we change optimization algorithms for this final state from the Residual Minimization Method with Direct Inversion in the Iterative Subspace (RMM-DIIS) to the conjugate gradient algorithm, which when combined with tight tolerances, yields higher confidence in the ground state configuration and energy.
These final runs reached the convergence criteria within one ionic step and only ran the minimum specified electronic steps, as the criteria was reached.
Figure~\ref{fig:equil} shows the total energy decreasing as a function of the ionic step in the minimization, reaching low values of total energy that satisfy the force tolerance and reaching a well-convergence ground state.

Figure~{\ref{fig:equil}a} shows the total energy of the system as a function of the ionic step for methane sI hydrates using the revPBE functional. 
This representative trend shows the rapid initial decrease followed by a slower approach towards the ground state.
While the total energy reached the minimum quickly, the force minimization is the objective in the ionic calculations.
Figure~{\ref{fig:equil}b} shows the total energy of the system for the carbon dioxide sI hydrate using the SCAN XC functional.
The decrease is indicative of the correct functioning of the calculation and, when combined with visual inspection of the atomic structure, confirms that the system has reached its ground state configuration, is stable, and has not broken apart or fracture.
Therefore, further analysis may proceed.

\subsection{Thermodynamic Backbone}

\begin{figure}
  \includegraphics[width=3.33in]{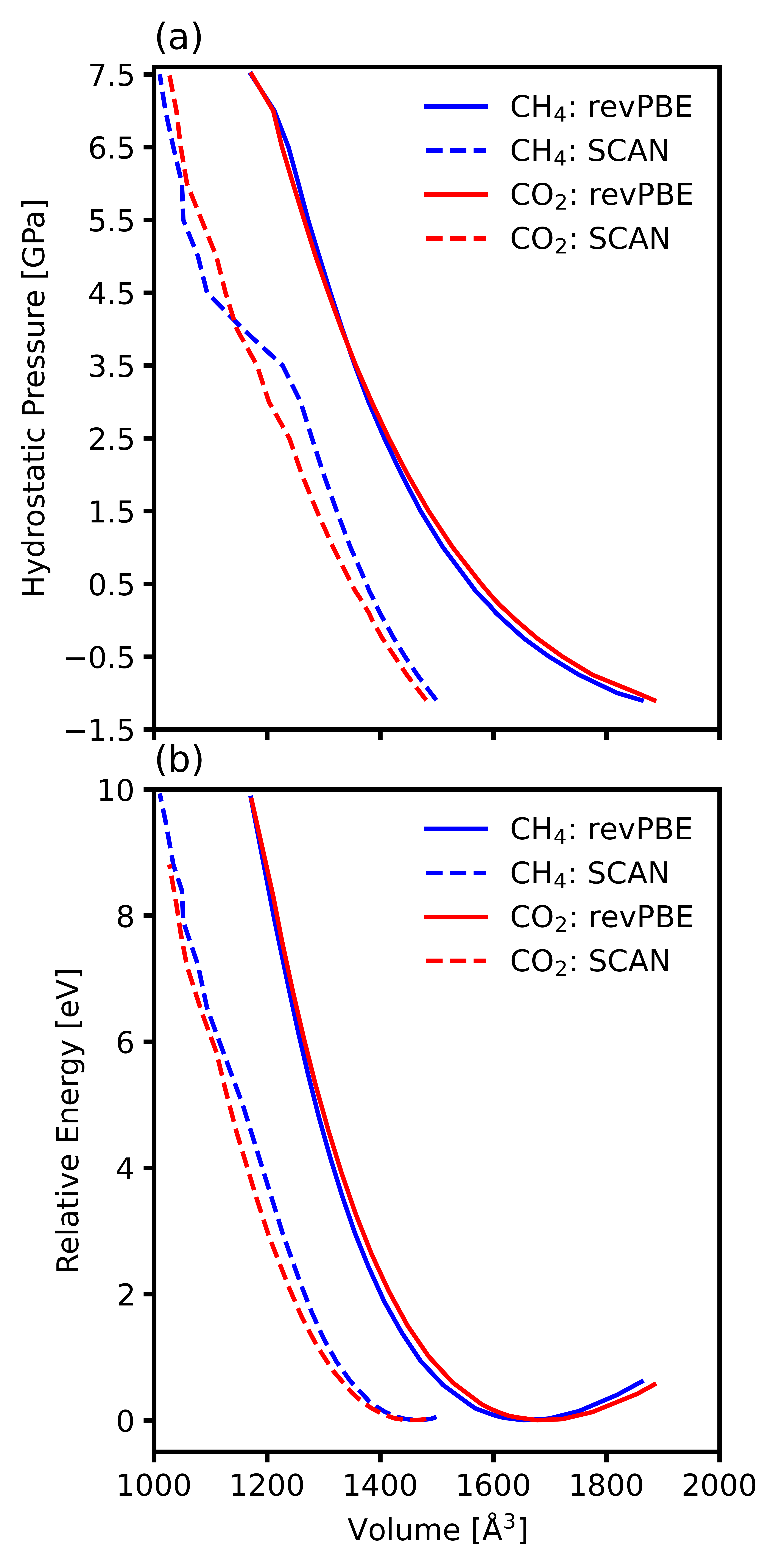}
  \caption{(a) The pressure-volume relationship of methane (blue) and carbon dioxide (red) sI gas hydrates simulated using the revPBE (solid line) and the SCAN (dashed line) XC functionals. (b) The energy-volume relationship of methane (blue) and carbon dioxide (red) sI gas hydrate simulated using the revPBE (solid line) and the SCAN (dashed line) XC functions.}
  \label{fig:pev}
\end{figure}

Before using the isothermal EOS to study the material properties of the structures, it is important to analyze the pressure and energy variance with volume to characterize the differences with guest molecule in these structures.
Additionally, this will elucidate any functional biases that exist in the computations between the revPBE and the SCAN XC functionals.
The unique hydrogen bond network of the hydrate structure is composed of a balance of VdW interactions, hydrogen bonds, and covalent bonds, and having a general purpose, nonempirical XC functional, like SCAN, provides accurate descriptions of this environment~\cite{Chen2017InitioTheoryModeling}.
Figure~\ref{fig:pev} shows relationship between the pressure, energy, and volume of the two systems generated with the two different functionals, providing visual proof of equilibrium volume differences.

Figure~{\ref{fig:pev}a} and~{\ref{fig:pev}b} show that the different guest molecules in structures treated with the revPBE functional possess little differences in trends and values across the pressure (and therefore volume) range.
In figure~{\ref{fig:pev}a}, revPBE provides no differentiation of the volume as a function of applied hydrostatic pressure.
Jendi~et~al. found that the revPBE functional also shows little difference between the two, with less than 0.2 \% difference in the equilibrium volume and 6 \% difference in the stiffness between carbon dioxide and methane sI hydrates~\cite{Jendi2014InitioDFTStudy}.
Figure~{\ref{fig:pev}b} displays the same behavior for energy, even at high compressions or tensions, nearing the stability limits.
In 100 \% occupied sII hydrates, simulation with revPBE gives differences of 0.7 \% and 0.15 \% for the equilibrium volume and stiffness, respectively~\cite{Vlasic2017EffectGuestSize}, for systems of methane-ethane and methane-propane, showing that differences in guest molecule size have relatively weak effect on the structure at low compression.

Figure~{\ref{fig:pev}} displays that the SCAN functional dramatically changes the equilibrium volume and the stiffness, to a lesser degree.
This XC functional corrects for the underbinding of revPBE, better describing the directional bonding, dispersion, and yielding a smaller equilibrium structure and a stiffer response to pressure.
Additionally, it captures the effect of different guest molecules on the backbone.
Figure~{\ref{fig:pev}a} shows that the methane sI hydrate simulation with the SCAN functional undergoes some structural rearrangemnet at pressures exceeding 3.5~GPa, indicated by the change in stiffness under further compression, but that the carbon dioxide system smoothly continues under the same regime.
These findings indicate the requirement for further analysis of the differences between the guest molecules and functionals.
Therefore, we compute the isothermal bulk modulus, the first pressure derivative of the bulk modulus, and the equilibrium volume of the sI stucture by fitting the energy volume curve from figure~{\ref{fig:pev}b} to the Vinet isothermal EOS, and compare existing literature values to ascertain the source of these differences.
The resulting parameters are presented in table~\ref{tab:eos}.
The stiffness and bulk modulus change with pressure is treated in a latter section.

\begin{table}
  \caption{Structure I gas hydrate properties with methane and carbon dioxide guest molecules calculated using different functionals.}
  \label{tab:eos}
  \begin{tabular}{llrrr}
    \hline
    \multicolumn{1}{c}{Guest} &
    \multicolumn{1}{c}{XC Functional} &
    \multicolumn{1}{c}{$B_{0}$ [GPa]} &
    \multicolumn{1}{c}{$B_{0}^{'}$} [$-$] &
    \multicolumn{1}{c}{$V_{0}$ [\AA$^{3}$]}  \\
    \hline
    methane\textsuperscript{\emph{a}} & revPBE & 10.58 & 6.08 & 1660.91 \\
    methane\textsuperscript{\emph{a}} & SCAN & 17.66 & 5.04 & 1465.06 \\
    carbon dioxide\textsuperscript{\emph{a}} & revPBE & 10.11 & 5.68 & 1687.83 \\
    carbon dioxide\textsuperscript{\emph{a}} & SCAN & 15.80 & 4.08 & 1456.43 \\
    methane\textsuperscript{\emph{b}} & revPBE & 10.69 & 6.41 & 1661.5\hphantom{0} \\
    methane\textsuperscript{\emph{c}} & revPBE & 9.98 & & 1786.45 \\
    methane\textsuperscript{\emph{d}} & BLYP & 8.3\hphantom{0} & & 1775.96 \\
    methane\textsuperscript{\emph{e}} & revPBE & 9.98 & 6.37 & 1684\hphantom{.00} \\
    carbon dioxide\textsuperscript{\emph{e}} & revPBE & 10.58 & 6.37 & 1687\hphantom{.00} \\
    \hline
  \end{tabular}

  \textsuperscript{\emph{a}}This work;
  \textsuperscript{\emph{b}}Zhu~et~al.~\cite{Zhu2022PiezoelasticityStabilityLimits};
  \textsuperscript{\emph{c}}Huo~et~al.~\cite{Huo2011MechanicalThermalProperties};
  \textsuperscript{\emph{d}}Miaranda~et~al.~\cite{Miranda2008FirstprinciplesStudyMechanical};
  \textsuperscript{\emph{e}}Jendi~et~al.~\cite{Jendi2014InitioDFTStudy}.
\end{table}

Compared with DFT results from literature, the results generated herein with the revPBE functional compare favorably, with low variance for the methane guest molecules in all three parameters of interest.
Generally speaking, our calculations for this functional neither over nor underestimate the parameters.
The Becke-Lee-Yang-Parr (BLYP) functional underestimates the bulk modulus for methane and overestimates the equilibrium volume.
Table~\ref{tab:eos} shows that the SCAN functional tends to underestimate the volumes and overestimate the cohesive energies, yielding lower equilibrium volumes and bulk moduli at zero pressure.
Additionally, the SCAN functional underestimates the pressure derivative of the bulk modulus.
The bulk modulus would be at its maximum at 0 K for a monocrystal and SCAN provides an upper limit in this regard~\cite{Douce2011ThermodynamicsEarthPlanets}.

\begin{figure}
  \includegraphics[width=6.5in]{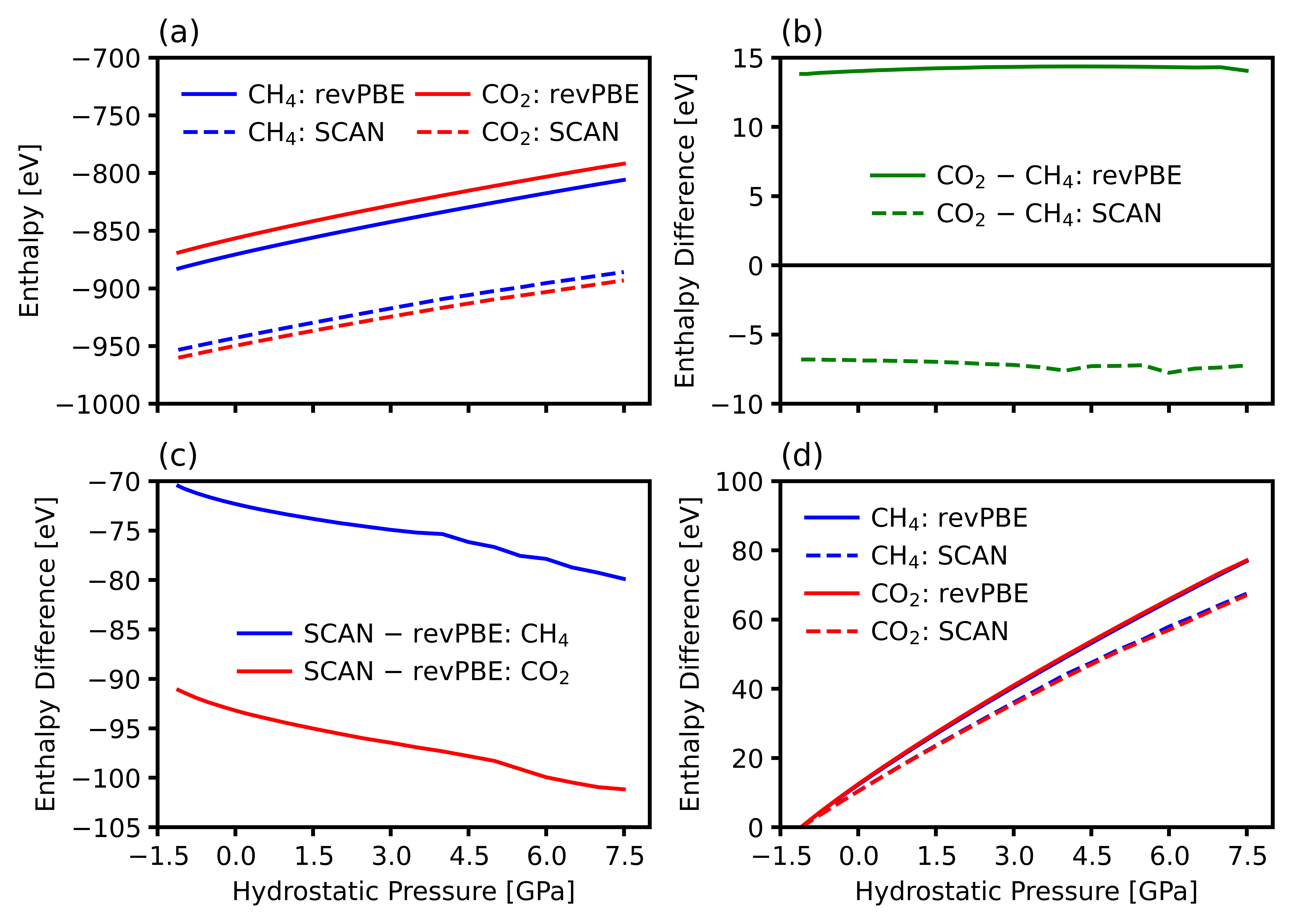}
  \caption{(a) The enthalpy of methane (blue) and carbon dioxide (red) sI gas hydrates simulated using the revPBE (solid line) and the SCAN (dashed line) XC functionals. (b) The enthalpy difference between carbon dioxide and methane hydrates simulated using the revPBE (solid line) and the SCAN (dashed line) XC functionals. (c) The enthalpy difference between the SCAN and revPBE XC functionals for sI methane (blue) and carbon dioxide (red) hydrates. (d) The enthalpy difference relative to the lowest enthalpy of methane (blue) and carbon dioxide (red) sI gas hydrates simulated using the revPBE (solid line) and the SCAN (dashed line) XC functionals. Note that (d) has four curves that are overlapping in pairs.}
  \label{fig:enthalpy}
\end{figure}

When comparing studies with theoretically more accurate dispersion corrections, sI methane hydrates were found to yield bulk moduli that were 68 \% higher, pressure derivatives that were 9.7 \% lower, and equilibrium volumes that were 12.3 \% lower~\cite{Jendi2014InitioDFTStudy}.
Our works finds the same trends, but with 67 \%, 17 \%, and 11.8 \% differences for sI methane and 56 \%, 28 \%, and 13.7 \% differences for sI carbon dioxide when comparing the revPBE and SCAN functionals, which both include dispersion corrections but, in the case of SCAN, more accurately.
This indicates that the inclusion of VdW and dispersion interactions, combined with the known underbinding of the revPBE XC functional, predicts stiffer materials with small equilibrium volumes.
The same behavior is also seen in sH hydrates systems with various guest gases~\cite{Daghash2019StructuralPropertiesSH}.
In our computations, the complexity of the long range bounding environment surrounding the carbon dioxide guest molecule causes differences in trends when comparing the effect of guest molecules, and this requires further analysis of stability with different criteria and examination of how these trends change under the application of hydrostatic pressure.
Therefore, we compute the enthalpy as of the system as a function of pressure to show any stability ordering and crossover pressures, and understand how pressure, combined with a specific guest molecule, affects the stability of the structure in figure~\ref{fig:enthalpy}.

The feasibility of a given structure's existence under various pressures is analyzed in terms of its enthalpy.
The enthalpy is a practical and critical thermodynamic property that provides information of the stability under particular states, and when the enthalpy difference is compared across different guest molecules and XC functionals, we can asses if there are any crossover points, which would indicate important structural changes in the system~\cite{Rychkov2017PressuredrivenPhaseTransition}
Additionally, it aids in analysis of stability under certain simulation criteria and to elucidate the source of property differences.
In figure~{\ref{fig:enthalpy}a}, the SCAN functional has a more negative enthalpy than the revPBE functional for both guest molecules, indicating that the SCAN provides a more stable simulated structure, meaning that under pressure it is able to distribute the stresses and energies better thanks to structural influences.

Figure~{\ref{fig:enthalpy}b} compares the enthalpy difference between the two guest molecules for a given functional.
We can see that the revPBE assesses the methane hydrate to have lower enthalpy across the entire pressure band, with no appreciable difference throughout the range.
Meanwhile, SCAN finds that the carbon dioxide hydrate has the lower enthalpy across the pressure range.
Thanks to its improved inclusion of medium to long range interactions, the SCAN functional can accurately simulate the complexity of the carbon dioxide's hydrogen bonding environment, clarifying the ability of the carbon dioxide molecule to possible rotate or rearrange itself and pack into the structure more efficiently.
Figure~{\ref{fig:enthalpy}c} shows that as pressure increases, the enthalpy difference caused by employing the SCAN versus the revPBE functional decreases, meaning that as the structure shrinks with increasing hydrostatic pressure, the medium to long range interactions captured by SCAN play an increasing role in stabilizing the structure as it deals with the same increasing pressure.
Finally, figure~{\ref{fig:enthalpy}d} shows that the XC functional plays a small role in differentiating the marginal enthalpy difference.
This is a clearer summary of figure~{\ref{fig:enthalpy}a}, where we see the lines are parallel by a constant offset.
This vertical offset means that since the two structures, simulated with a given functional, have the same pressure, the guest is reducing the enthalpy of the structure by a constant amount over the pressure window.

\subsection{Mechanical Stability}

\begin{figure}
  \includegraphics[width=3.33in]{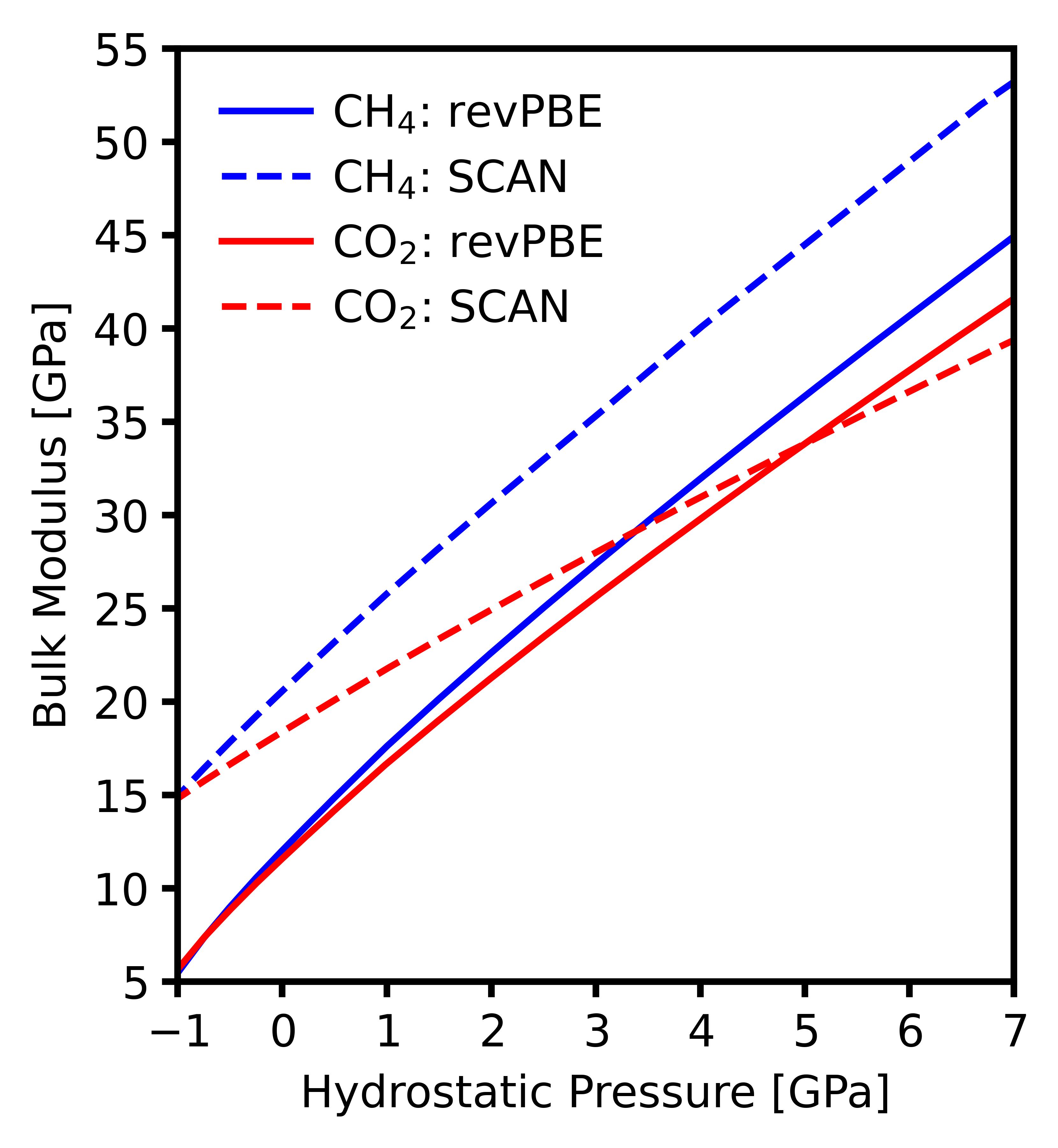}
  \caption{Bulk modulus as a function of pressure of methane (blue) and carbon dioxide (red) sI gas hydrates simulated using the revPBE (solid line) and the SCAN (dashed line) XC functionals.}
  \label{fig:modulus}
\end{figure}

Figure~\ref{fig:modulus} shows the bulk modulus as a function of pressure for the systems involved calculated using the Vinet EOS.
The bulk modulus shows an increasing trends throughout the pressure range, from $-1.1$ GPa to $7.0$ GPa. 
Both guest molecules simulated with both functionals show the trend for resistance to uniform compression.
The revPBE functional yields only a small difference that increases with pressure, and the carbon dioxide sI hydrate is actually less resistant to uniform compression.
The SCAN functional shows a much larger difference as the pressure is increased.
The difference between the methane and carbon dioxide bulk modulus increases with pressure, indicating the level of resistance within the structure is lower and a higher compressibility.
Due to the heightened interaction between the carbon dioxide and the backbone, which is only increased as the structure shrinks.
The increased electron density of the oxygen atoms in carbon dioxide mimics an increasing guest molecule, causing some repulsive interactions.
In sII gas hydrates, larger guest molecules caused lower bulk moduli as well~\cite{Vlasic2017EffectGuestSize}.

The bulk modulus quantifies a material's resistance to uniform hydrostatic compression, and is the product of the volume and the deriative of the pressure with respect to the volume.
For molecular crystals and hydrates, it typically increases with pressure since short-range repulsive interactions are enhanced, compressibility is reduced, and intermolecular distances shrink.
The stiffening is well captured by the Vinet EOS, showing a positive pressure derivative of the bulk modulus.
The hydrogen bonding and cage geometry in gas hydrates dominate the compressional regime~\cite{Ning2015CompressibilityThermalExpansion}.
As the SCAN functional yields a small eqilibrium volume, this artificial compression, compared to the revPBE, gives larger values to the bulk modulus while simulateneously resolving interactions that produce more complex and differing behaviors with increasing pressure.

Figure~\ref{fig:modulus} shows that the choice of XC functional strongly influences the predicted stiffness.
The family of meta-GGA functionals, of which SCAN is a member, generally yield higher bulk moduli for hydrate materials than normal GGA functionals, like revPBE, because they enforce exact constraints and include the kinetic energy, improving the description of hydrogen bonding and intermediate-range interactions like VdW and dispersion~\cite{Kovacs2019ComparativeStudyPBE}.
In turn, this gives rise to stronger guest-host coupling and smaller equilibrium values.
Our results show that revPBe tends to underbind, giving similar bulk moduli for the two guest molecules.
In contrast, SCAN amplifies the differences between the two guests.
Carbon dioxide is a linear molecular and more polarizable, and therefore interacts more strongly with the water lattice when simulated with the SCAN functional, causing cage distortion and lower stiffness.

The complexities of structural stiffness, bulk modulus, enthalpy differences, and equilibrium properties gives rise to a desire to understand the atomic origins of these behaviors.
Our previous work on the fracture mechanics of methane hydrates showed that various pathways exist for these materials fail~\cite{Zhu2023AtomisticgeometricSimulationsInvestigate,Zhu2023DFTcontinuumCharacterizationThirdorder}.
Additionally, we found that a geometric analysis of the backbone structure provides an accurate way to predict which parts of the structure will undergo failure first, even before complete structural breakdown~\cite{Mathews2025GeometricCharacterizationsNonUniform}.
Therefore, to compare the effect of guest molecules and their treatment by different XC functionals, we proceed with comparing the behaviors of the guest molecules in the structure under hydrostatic pressure to clarify how methane or carbon dioxide, under simulation with revPBE and SCAN, evolve and move to participate in the overall system response.

\subsection{Structural Evolution}

\begin{figure}
  \includegraphics[width=6.5in]{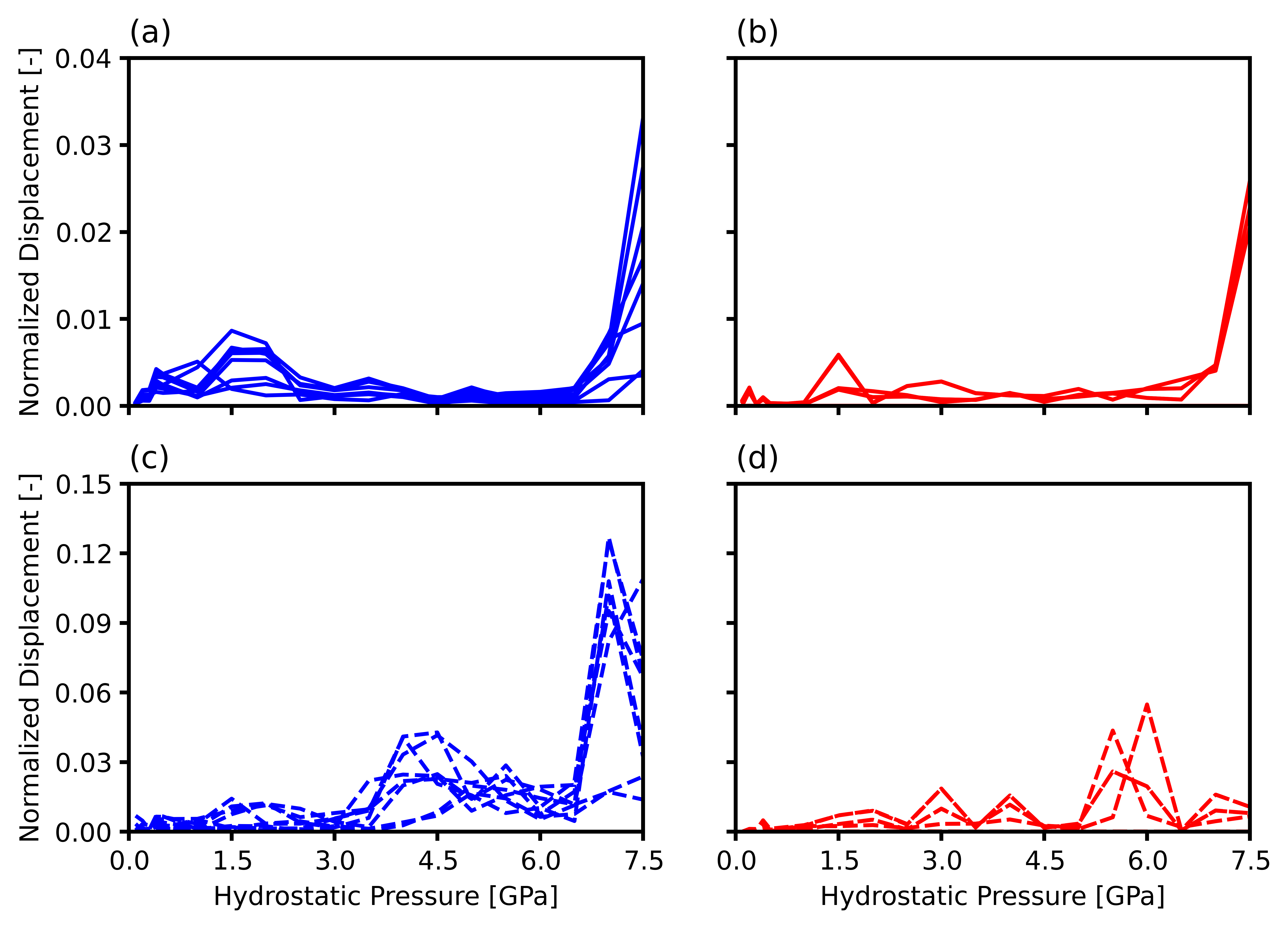}
  \caption{Normalized displacement of guest molcules in sI gas hydrates under compression: (a) Methane with revPBE, (b) Carbon dioxide with revPBE, (c) Methane with SCAN, (d) Carbon dioxide with SCAN.}
  \label{fig:displacement}
\end{figure}

Previous studies in gas hydrate stability focused on differences in guest-host interactions or cohesive energy, and did not adequately study the common characteristics of the these interactions~\cite{Chen2021ExploringGuestHost}.
Therefore, we first seek to analyze the displacement of the guest molecule within the hydrate cages.
Within these cages, the guest molecules remain confined within their cages even at higher pressures.
Experimental and simulation studies show that increasing pressure reduces the free volume in the cages, restricting translational and rotational motion of the guest special, leading to orientational ordering rather than large-scale displacements, a key factor in mechanical resilience of hydrates, given that the guest molecules are stabilizers during compression~\cite{Hirai2023SignificanceHighpressureProperties,Mathews2025GeometricCharacterizationsNonUniform}.
Therefore, we begin by assessing the displacement of the guest molecules under compression for revPBE and the SCAN functional to understand how the XC functional changes these behaviors in figure~\ref{fig:displacement}.

In Figure~\ref{fig:displacement}, the normalized displacement is calculated based on positions calculated from the evolving lattice parameters, which decrease under pressure.
Using this method extricates the displacement caused by the distances between all the molecules shrinking under pressure, producing a movement only coming from the change in the guest molecule as it responds to the shrinking cages.
The displacement is based on the position of the central carbon atom.
Figure~{\ref{fig:displacement}a} and~{\ref{fig:displacement}b} show that the carbon dioxide guest molecule have lower variability in their movement around the cages when simulated with revPBE, while methane molecules move more, particularly during initial compression.
Figure~{\ref{fig:displacement}c} and~{\ref{fig:displacement}d} show the same phenomenon, albeit the amplitude of the normalized displacement is higher.
The SCAN meta-GGA functional captures intermediate-range interactions and hydrogen bonding more accurately than revBE, causing higher variability in normalized displacement. 
Its ability to capture more nuances of the complicated energy landscape in hydrate cages across the pressure range at various interaction distances leads to this behavior.
The simplest representation of the energy landscape as a purely distance based surface, where different atoms, molecules, masses, and bonds are not resolved, leads to the simplest energy landscape that is pressure invariate~\cite{Mathews2024ModelingEffectBackbone}.

\begin{figure}
  \includegraphics[width=6.5in]{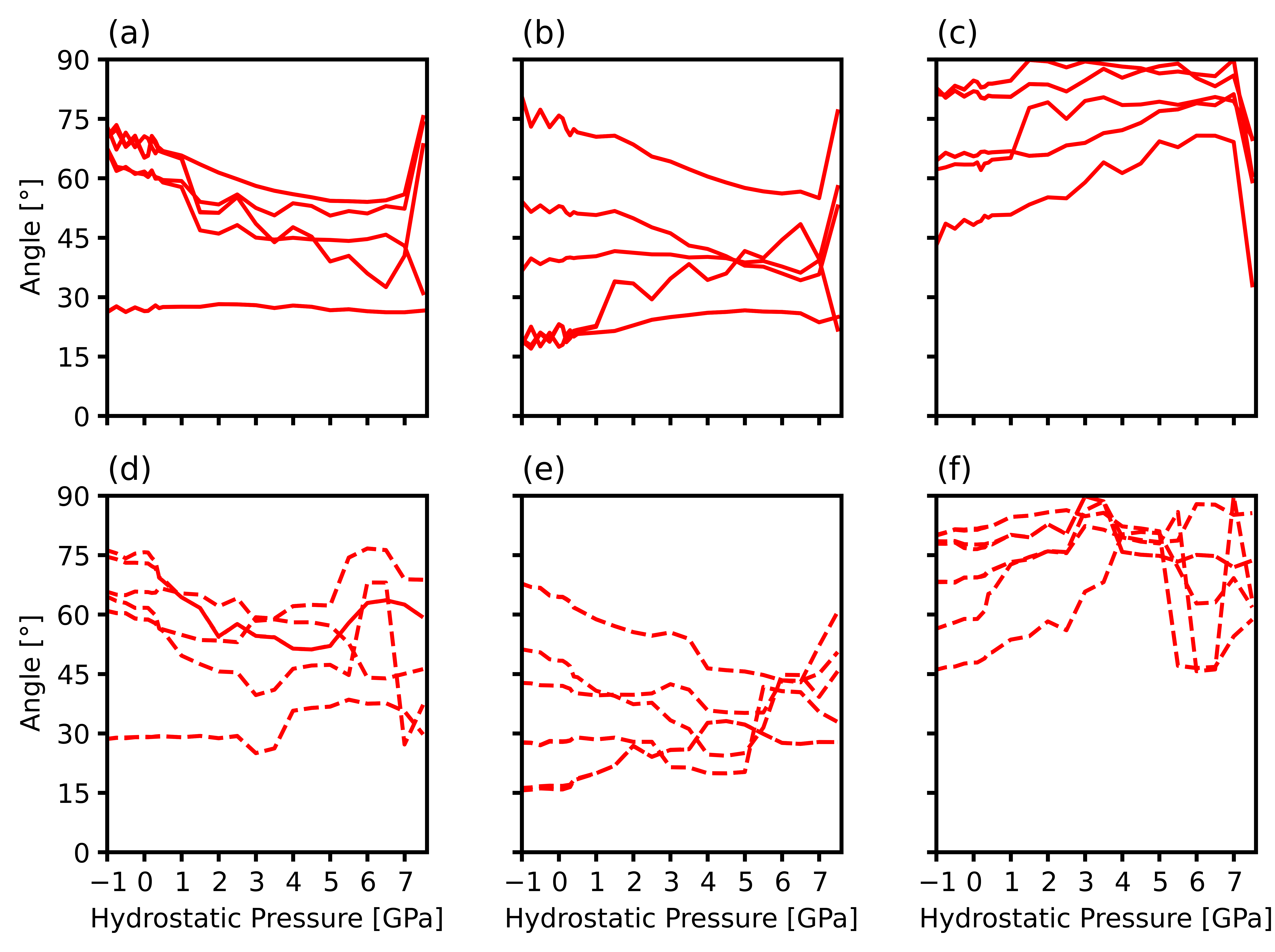}
  \caption{The angle between the carbon dioxide molecule and the three lattice parameters as a function of pressure. (a), (b), and (c) are the angles simulated with revPBE. (d), (e), and (f) are the angles simulated with SCAN. (a) and (d), (b) and (e), (c) and (f) represent the same lattice parameters.}
  \label{fig:rotation}
\end{figure}

Methane molecules in sI hydrates are smaller and relatively spherical, meaning that they interact more weakly with the lattice than the carbon dioxide molecule, meaning the energy landscale is relatively flat.
Its lack of strong hydrogen bonding makes positional fluctuations energetically inexpensive.
The carbon dioxide molecular, by contrast, has a quadrupole moment that creates directional interactions in the network will provide another way for the molecular to interact with the backbone.
As the pressure increases, the molecule compensates for the lack of displacement through rotational adjustments.
We quantify these rotational adjustments in figure~\ref{fig:rotation}.

In figure~\ref{fig:rotation}, the angles are computed as the angles between 3D vectors representing each of the eight carbon dioxide guest molecules and the three lattice parameters.
While as zero pressure the lattice parameters are all orthogonal, the carbon dioxide molecule shows some alignment that affects the unit cell's response to hydrostatic pressure, causing the unit cell to lose its pure cubic structure~\cite{Izquierdo-Ruiz2016EffectsCO2Guest,Mathews2020HeatCapacityThermal}.
Therefore, we treat the angles and lattice parameters separately.
The rotation of the methane molecules is not studied here as they are able to rotate freely accoring to simulation~\cite{Izquierdo-Ruiz2015GuesthostInteractionsGas} and experiment~\cite{Takeuchi2013WaterProtonConfigurations}.

Figure~{\ref{fig:rotation}a},~{\ref{fig:rotation}b}, and~{\ref{fig:rotation}c} show rotation towards specific orientations for all eight guest molecules, as evidenced the angles converging towards similar values with increasing pressure, especially for~{\ref{fig:rotation}c}.
In~{\ref{fig:rotation}c}, as the pressure increases, the approach towards 90 degrees corresponds to alignment parallel to the hexagonal phases of the large cages of structure ($5^{12}6^{2}$).
This alignment corresponds to experimental diffraction data that showed that the most frequent orientation of the guest molecule is not totally parallel to the hexagonal face but instead tilted by between 6.5 and 14.4 degrees~\cite{Udachin2001StructureCompositionThermal}, while our work shows a spread of 20 degrees under the highest pressure, just before fracture and decomposition evidenced by the dramatic change in orientation.

Figure~{\ref{fig:rotation}d},~{\ref{fig:rotation}e}, and~{\ref{fig:rotation}d} show more variability as the pressure is increased because of the same aforementioned reshaping of the energy landscape caused by improved resolution of the SCAN functional.
We see some similar alignment and collapse of the distribution, but less so when compared with revPBE, because of the treatment of SCAN of more complex interactions.
While theoretical studies of hydrates under pressure focusing on the water molecules considered the backbone and its changes under pressure with only revPBE~\cite{Mathews2025GeometricCharacterizationsNonUniform,Zhu2022MultiscalePiezoelasticityMethane,JafariDaghalianSofla2024AtomisticGeometricContinuum}, in this work we focus on the methane and carbon dioxide, simulated with revPBE and SCAN XC functionals to elucidate the role they play in stabilizing the structure under pressure.
As such, the findings presented here provide a method of identifying where the XC functional participated in the material properties of gas hydrates and using first principles calculations to explain the orientational distribution of movement of guest molecules, while expaining differences in marginal stability that are produced by using different functionals.

\section{Conclusions}

The pressure-enthalpy landscape and mechanical stability envelope of sI gas hydrates are governed by guest-host sterics and electrostatics in a way that produces signatures in equation of state parameters, cage level distortions, and guest orientations.
These signatures can aid in prediction of the relative stability of methane versus carbon dioxide hydrates, and their hydrostatic strength limits under compression.
The SCAN + rVV10 XC functional captures the trends more finely than revPBE + DFT-D2, but the structure-property map is consistent across the functionals.
This work investigated sI methane and carbon dioxide hydrates using the revPBE + DFT-D2 and SCAN + rVV10 XC functionals through computational materials science methods based on density functional theory to assess the differences in material properties and their atomic cause.
This finding validates some anomoulous compressibility behaviors of guest molecules with strong quadrupole moments like carbon dioxide and ethylene oxide~\cite{Mathews2020HeatCapacityThermal}.

The results herein serve to fill the gap in current knowledge regarding the molecular, orientational, and material properties of sI methane and carbon dioxide hydrates simulated with the commonly used revPBE + DFT-D2 XC functional, and supplement these findings by comparing them with findings from using the modern SCAN + rVV10 XC functional.
The theoretical findings show that the guest contributes importantly in the material properties, like bulk modulus, and determine the marginal stability of the different sI hydrate systems.
The specific findings in enthalpy differences, bulk modulus, molecular displacement, and guest molecule orientation can be applied to any inclusion compound where differences and stabilization provided by the guest molecule remains elusive.
Additionally, the features herein can be extracted from existing characterisation methods to help extend understanding material properties across pressure conditions beyond those at which they were obtained. 
Following this multiparameter methodology, the adaptive geometry of crystalline sI gas hydrates is decribed and defined beyond only considering these structures as a backbone capturing distinct guest molecules, but as solid phases of lattices interacting strongly with the guest molecules.

Our first-principles simulations provide novel confirmation of experimental observations that the carbon dioxide guest molecule exhibits restricted translational motion in the cages of sI hydrates.
Unlike methane molecules, it maintains the center-of-mass positioning but compensates through rotational adjustments.
This directional behavior, captured with the advanced SCAN XC functional, underscores strong coupling between the linear molecule's polarity and the hydrogen bond network.
These findings bridge the gap between computational and experimental perspectives, offering insight what governs hydrate stability in extreme conditions.

\begin{acknowledgement}
  This work was supported by the Fonds de Recherche du Québec Nature et Technologies through the Bourse de Doctorat en Recherche.
  This research was enabled, in part, by support provided by Calcul Quebec, the BC DRI Group, and the Digital Research Alliance of Canada (\href{https://alliancecan.ca/}{alliancecan.ca}).
\end{acknowledgement}

%

\bibliography{References}

\end{document}